\documentclass[twocolumn,showpacs,preprintnumbers,amsmath,amssymb]{revtex4}

\usepackage{graphicx}
\usepackage{dcolumn}
\usepackage{bm}

\begin{document}


\title{New combined PIC-MCC approach for 
fast simulation of a radio frequency 
discharge at low gas pressure.}
\author{I.V.~Schweigert}%
\email{ischweig@itam.nsc.ru}
\affiliation{Institute of Theoretical and Applied Mechanics,
630090 Novosibirsk, Russia}
\author{V.A.~Schweigert}{
\affiliation{Institute of Theoretical and Applied Mechanics,
630090 Novosibirsk, Russia}
\date{\today}

\begin{abstract}
A new combined  PIC-MCC approach is developed for 
accurate and fast simulation of a radio frequency 
discharge at low gas pressure and high density of plasma. 
Test calculations of transition between different 
modes of electron heating 
in a ccrf discharge in helium and argon show 
a good agreement with experimental data.
 We demonstrate high 
efficiency of the combined PIC-MCC algorithm, 
especially for the collisionless regime of  electron heating.
\end{abstract}

\date{\today}

\maketitle

\section {Introduction}

The modern
trends of plasma technologies  are directed to a reduction of 
the gas
pressure and an increase of plasma density. The further 
development of efficient
methods is required for simulation of collisionless regimes in a 
capacitevely coupled and especially in an inductively coupled 
discharges as the collisionless heating of  electrons plays a 
key role in dynamics of  thin skin layers.
The Particle-in-Cell Monte Carlo Collisions (PIC-MCC) method 
\cite{bird} has
become a standard simulation technique for a gas discharge in 
plasma reactors
of etching or deposition. Unlike the fluid approach, the PIC-MCC 
algorithm requires
larger computer resources, but it provides a detail kinetic 
picture of processes in a gas discharge.
However, a problem 
of statistical fluctuations of
an electric field appears at low gas pressures, in particular 
for gases with a deep Ramsauer minimum
in the elastic scattering cross section. 
For the periodic electrical field $%
E=E_0\sin(\omega t)$, where $\omega$ is the discharge frequency, 
the rate of electron heating in the bulk is
proportional to $\nu E_0^2/(\omega^2+\nu^2)$, 
where $\nu $ is electron collision
frequency.  At high gas pressure 
in the
collisional regime, when $\nu >\omega$,
the electric field in the quasineutral plasma
is sufficiently large and the effect the artificial electron 
heating 
is less dangerous. At the low gas pressure in the collisionless 
regime (at $%
\nu<\omega$), the electrons gain the energy in the electrode 
sheaths. In the quasineutral plasma the
electric field is small and the 
 electrons scattering on the field fluctuations  essentially 
distorts
the results. Although the numerical smoothing of the charge 
density \cite{bird2} helps 
to diminish the statistical noise, it is necessary to develop a 
more
radical way for reduction of the influence of statistical 
fluctuations.

An interesting idea was suggested in Ref.\cite{ivan}. As the
discharge simulation lasts more than one thousand of  discharge periods, 
the averaging of the charge density over
several periods 
reduces the statistical noise. But the direct 
averaging can lead to the  development of the numerical 
instability. To eliminate this problem, the electric field was 
calculated
in Ref.\cite{ivan} from the current continuity equation. 
However, this approach requires an explicit
distinction of electrode sheaths, that is difficult  for 
realization in the 
two-dimensional case. Besides, it does not take into account  
inertia of
electrons, which is very important at the low gas pressure. 
Below we present
 another way of the noise reduction in a new approach developed 
by Vitaly Schweigert.

\section {Combined PIC-MCC approach}

In the combined PIC-MCC approach we find the electric field 
distribution from the auxiliary equations which are derived from 
the kinetic equations. The integration of the electron and ion 
kinetic equations over the velocity gives us the continuity
equations for electron and ion densities.
The integration of the kinetic equations multiplied by the 
velocity gives the continuity equations for electron and
ion fluxes. The kinetic coefficients are calculated with using 
the
electron and ion distribution functions, which are found from 
the electron and ion kinetic equations. 
 To avoid the kinetic
coefficients fluctuations we average them over many periods.  
Thus, in our model the kinetic equations, the auxiliary 
equations and the Poisson equation are solved self-consistently. 
The kinetic approach allows us to find the kinetic
coefficients and  the electric
field distribution is found from the auxiliary equation. 
The equation system includes
the Boltzmann kinetic equations  for velocity 
distribution 
functions of electrons $f_e(t,x,\vec v)$ 
and
ions $f_i(t,x,\vec v)$, which are three dimensional over the 
velocity and one dimensional in the space
\begin{equation}  \label{kine}
\frac {\partial f_e}{\partial t}+ \vec v_e\frac {\partial 
f_e}{\partial x}
-\frac {e\vec E}{m}\frac {\partial f_e}{\partial \vec v_e}= 
J_e,\quad n_e=\int
f_ed\vec v_e,
\end{equation}
\begin{equation}  \label{kini}
\frac {\partial f_i}{\partial t}+ \vec v_i\frac {\partial 
f_i}{\partial x}
+ \frac {e\vec E}{M}\frac {\partial f_i}{\partial \vec v_i} 
=J_i,\quad
n_i=\int f_id\vec v_i,
\end{equation}
where $v_e$, $v_i$, $n_e$, $n_i$, $m$, $M$ are the electron and 
ion 
velocities, densities and masses, respectively, $\vec E$ is the 
electrical field, $J_e$, $J_i$ are the collisional integrals
for electrons and ions,  
the transport equations for the density and the flux of 
electrons and ions based on the
momentum of the kinetic equations 
(\ref{kine}),(\ref{kini})
\begin{equation}  \label{eqn1}
\frac{\partial n_e^{\prime}}{\partial t}+\frac{\partial 
j_e^{\prime}}{%
\partial x}=Q,
\end{equation}
\begin{equation}  \label{eqn2}
\frac{\partial n_i^{\prime}}{\partial t}+\frac{\partial 
j_i^{\prime}}{%
\partial x}=Q,
\end{equation}
\begin{equation}  \label{eqn3}
\frac{\partial j_e^{\prime}}{\partial t}=-\frac{\partial
T_e^{\prime}n_e^{\prime}}{\partial x} -\frac{eE}{m}n_e^{\prime}-
\nu_ej_e^{%
\prime}-Q_e,
\end{equation}
\begin{equation}  \label{eqn4}
\frac{\partial j_i^{\prime}}{\partial t}=-\frac{\partial
T_i^{\prime}n_i^{\prime}}{\partial x} +\frac{eE}{M}n_i^{\prime}-
\nu_ij_i^{%
\prime}-Q_i,
\end{equation}
where 
\begin{equation}  \label{dum1}
Q=N_g\int v\sigma_if_ed\vec v_e
\end{equation}
is the ionization rate, $\sigma_i$ is the ionization cross 
sections, $N_g$
is the gas density, 
\begin{equation}  \label{dum2}
T_e^{\prime}=\frac{\int v_{ex}^2f_ed\vec v_e}{\int f_ed\vec 
v_e},\quad T_i^{\prime}=%
\frac{\int v_{ix}^2f_id\vec v_i}{\int f_id\vec v_i}
\end{equation}
are the effective electron and ion temperatures, respectively, 
$$Q_e=N_g\int v_{ex}|\vec v_e|\sigma_tf_ed\vec v_e-\nu_e\int 
v_{ex}f_ed\vec v_e,\quad$$
$$Q_i=N_g\int v_{ix}|\vec v_i|\sigma_rf_id\vec v_i-\nu_i\int 
v_{ix}f_id\vec v_i$$
describe the friction for electrons and ions, the efficient 
frequencies 
\begin{equation}  \label{dum4}
\nu_e=\frac{N_g\int |\vec v_e|\sigma_tf_ed\vec v_e}{\int 
f_ed\vec v_e}, \quad\nu_i=%
\frac{N_g\int |\vec v_i|\sigma_rf_id\vec v_i}{\int f_id\vec 
v_i},
\end{equation}
where $\sigma_t$ is the electron
transport cross section, $\sigma_r$ is the ion resonance charge 
exchange cross
section. Notice that in the usual fluid approach the terms 
$Q_e$, $Q_i$ are
supposed to be zero, which is correct only for the constant 
scattering frequencies.
 The boundary conditions for the auxiliary equations includes 
the secondary emission as in Ref.\cite {Boeuf1987}.
It can be easily seen, that the equations (\ref{eqn1})
-(\ref{eqn4}) are direct
consequences of the kinetic equations (\ref{kine}),(\ref{kini}). 
As we calculate
the kinetic coefficients $Q$, $T_e^{\prime}$, $T_i^{\prime}$, 
$Q_e$, $Q_i$
with solving the kinetic equations (\ref{kine}),(\ref{kini}) 
with the
Monte-Carlo method, the obtained densities $n_e^{\prime}$, 
$n_i^{\prime}$
have to coincide with a good accuracy with values from the 
kinetic equations
(\ref{kine}),(\ref{kini}). After calculating the auxiliary 
values of
electron $n_e^{\prime}$ and ion $n_i^{\prime}$ densities, we 
calculate the electric field from the Poisson equation 
\begin{equation}  \label{eqn5}
\bigtriangleup \phi =4\pi e (n_e^{\prime}-n_i^{\prime}), \quad 
E=-\frac{%
\partial \phi}{\partial x}.
\end{equation}
The reduction of statistical noise in our  approach is
reached with averaging the kinetic coefficients $Q$, 
$T_e^{\prime}$, $%
T_i^{\prime}$, $Q_e$, $Q_i$ over many periods and with  
smoothing over the 
spatial coordinate. For averaging a function $F(x)$ over 
preceding periods
we use the following algorithm 
\begin{equation}  \label{dum5}
F(x)^i=\alpha F(x)^{\prime i}+(1-\alpha)F(x)^{i-1},
\end{equation}
where $F(x)^{\prime i}$ is the value on the $i$-period  and $%
\alpha=0.01\div 0.1$. The spatial smoothing is chosen as in 
Ref.\cite
{bird2} 
\begin{equation}  \label{dum6}
F(x_k)=\frac{F(x_{k+1})+2F(x_k)+F(x_{k-1})}{4},
\end{equation}
where $x_k$ is the node of the simulation grid. 
The spatial smoothing is very important for resolving the space 
charge 
in the quasineutral part of a discharge, where the charge is a 
small difference of two large 
and almost equal values (ion and electron densities). 
The computer resources 
for solving the transport 
equations (\ref{eqn1})-(\ref{eqn4}) are much smaller 
than for the kinetic equations, 
therefore the auxiliary equations are solved for each
period, 
and 
the kinetic coefficients are calculated after several periods 
from the kinetic equations 
(\ref{kine}),(\ref{kini}). Then
the electron and ion
weights are fitted with the densities $n_e^{\prime}$, 
$n_i^{\prime}$.
We use $5000$ simulation particles for each charged species, the 
Cloud-in-Cell charge assignment
scheme, the null-collisions technique to find the time of  
electron and ion free 
flight, and
the energy conserving scheme with a second order of accuracy to 
solve the
equations of motion \cite{bird2,hock}.
The equations (\ref{eqn1})-(\ref{eqn4}) are solved with an 
implicit
finite-difference method with using the Scharfetter and Gummel 
scheme 
\cite{gummel}. 
 For small grid spacing $T_e\gg e|\phi_{k+1}-\phi_k|$ this
finite--difference scheme has a second order accuracy in 
$\bigtriangleup x$
and gives a correct result on rough grids for the Boltzmann 
electron
distribution. Like for the explicit PIC-MCC method, there
exists a restriction on time step $\omega_p\bigtriangleup t<1$, 
where $\omega_p$ is the plasma frequency, for solution
of the equations (\ref{eqn1})-(\ref{eqn4}) with the Poisson 
equation (\ref{eqn5}).

Since the
cross section of the electron Coulomb scattering is proportional 
to the electron density
and inversely proportional to the square of the electron energy, 
a correct
discharge simulation of some regimes requires accounting for 
electron Coulomb collisions.
For description of these collisions we apply the method 
\cite{manh}, where
the Langevin force and friction of electrons are introduced and 
defined from
their distribution function. 
Note also, that the Coulomb collisions do not change the total 
electron momentum of motion.

The test calculations show that this algorithm is 
numerically stable and allows one to reach a significant 
acceleration of the
PIC-MCC method due to two factors. At first, the time step for 
solving the
kinetic equations (\ref{kine}),(\ref{kini}) 
with implicit scheme 
does not depend on the plasma
frequency \cite{bird2,bird1}. 
At second, averaging
over many periods allows one to reduce greatly (in 5-20 times) 
the total
number of simulation particles in the PIC-MCC method without an 
increase of the statistical noise.

\section {How many simulation particles we need?}
We study a
capacitively coupled radio frequency (ccrf)
 discharge in argon and helium with the combined 
PIC-MCC approach for the experimental 
conditions of Godyak {\it et al}\cite{Godyak}.
We consider an
one-dimensional symmetrical ccrf discharge at room temperature 
for
the frequency $\nu =13.56~MHz$ and with the
sinusoidal shape of the discharge current $j$. 
One electrode is grounded and 
the voltage on another electrode is calculated self-consistently 
to provide the desired amplitude of the discharge current.
The spatial grid has typically $81$ nodes, condensing
in electrode sheaths. The minimal grid spacing
is decreased with the gas pressure rise, thus the sheath 
contains
approximately the constant number of nodes.  
The cross sections of electron scattering in helium are
taken from \cite{lagu}, and for argon from \cite{ivan,lagu}. The
ion--electron emission on electrodes is taken into account with 
coefficient 
$0.2$ in helium and $0.1$ in argon.

It is  known that the 
statistical error of Monte-Carlo methods
decreases as $1/N^2$. The statistical noise leads to 
the systematical error in
the electron cooling or heating. 
Therefore, first we have studied the impact of the number of 
simulation particles on an accuracy of results in three 
different methods: in the standard PIC-MCC 
 \cite {bird}, in the PIC-MCC with the spatial 
smoothing (PIC-MCC SS) \cite 
{bird2} and in our combined PIC-MCC.
 The simulations are performed for two values of argon pressures 
$P=0.1,~0.3$~Torr, 
 the inter-electrode distance $d=2$~cm and the discharge current 
$j=2.65$~mA/cm$^2$.  
The mean electron energy in the discharge center calculated with 
three methods and measured in Ref.\cite {Godyak} is shown in 
Fig.~\ref{comparison}.
\begin{figure}[tbp]
\resizebox{.8\columnwidth}{!}
  {\includegraphics[draft=false]{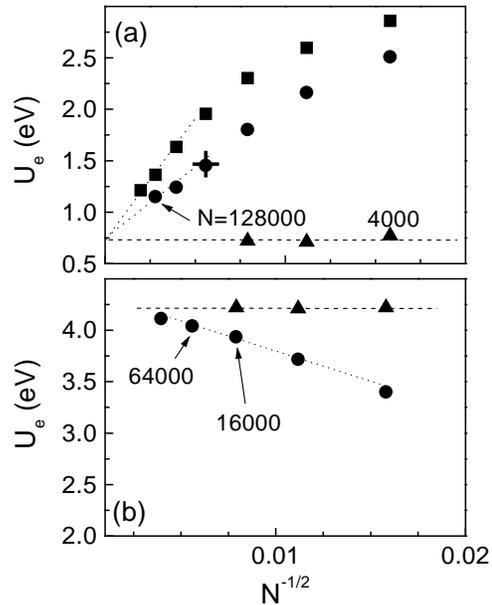}}
\caption {
Mean electron energy in the discharge center as a function of 
the total number of simulation particles for $P=0.1$~Torr (a) 
and $P=0.3$~Torr (b) calculated with the standard PIC-MCC method 
(squares), with the PIC-MCC SS method (circles) with spatial 
smoothing of the space charge and electrical field distributions 
and with our new combined algorithm (triangles).  
'Cross' is calculation from Ref.\protect\cite {bird2} 
with $N=32000$. $d=2$~cm, 
$j=2.65$~mA/cm$^2$. 
}
\label
{comparison}
\end{figure}
\begin{figure}[tbp]
\resizebox{.7\columnwidth}{!}
  {\includegraphics[draft=false]{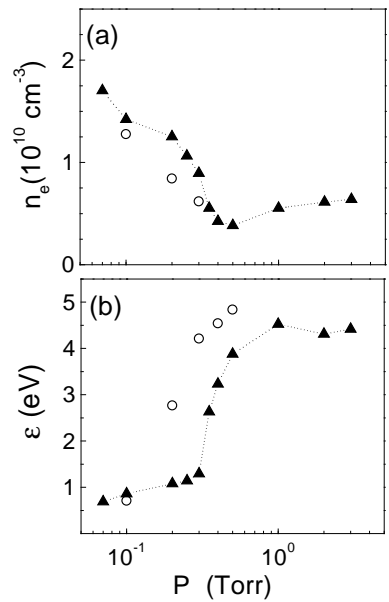}}
\caption {
Electron density (a) and  mean electron energy  (b) in the
discharge center ($x=1$~cm)  in argon computed 
(circles) and 
measured in \protect\cite {Godyak} (triangles) for
$d=2$~cm, $j=2.65$~mA/cm$^2$ and  $N=5000$.
}
\label
{density}
\end{figure}
The calculations with the standard PIC-MCC method with different 
$N$ 
show a significant role of electric field
fluctuations under the lower gas pressure (squares in  
Fig.~\ref{comparison}(a)).
It is seen that the standard PIC-MCC considerably overestimates 
the value of $\epsilon $ for $N=4000\div 256000$. 
The second method (PIC-MCC SS) gives much better results 
(circles  in Fig.~\ref{comparison}). 
The spatial smoothing indeed  
 decreases
the statistical noise, but feasibility of this technique is
restricted, since it distorts the space charge in the
electrode sheath.
At gas pressure $P=0.3~$Torr (see, Fig.~\ref{comparison}(b)) 
when the electron energy $\epsilon$ increases with $N$, 
the PIC-MCC SS method shows 
the reasonable accuracy (within $10\%$)
with small number of simulation particles $N=10000$. But
at the lower pressure $P=0.1$~Torr, the PIC-MCC SS 
is not able to provide convergency
in the electron energy even with $N=256000$.
It is obvious that at low gas pressure in order 
to obtain the reasonable solution with the standard PIC-MCC 
methods we need so an enormous number of the simulation 
particles that these methods are not more applicable.
As seen in Fig.~\ref {comparison} our combined PIC-MCC method 
gives the electron energy which is very close to the 
experimental one (see, Fig.~\ref {density})
 already with the small number of simulation particles and the 
results only weakly 
depend on $N$. 
\begin{figure}[tbp]
\resizebox{1.\columnwidth}{!}
  {\includegraphics[draft=false]{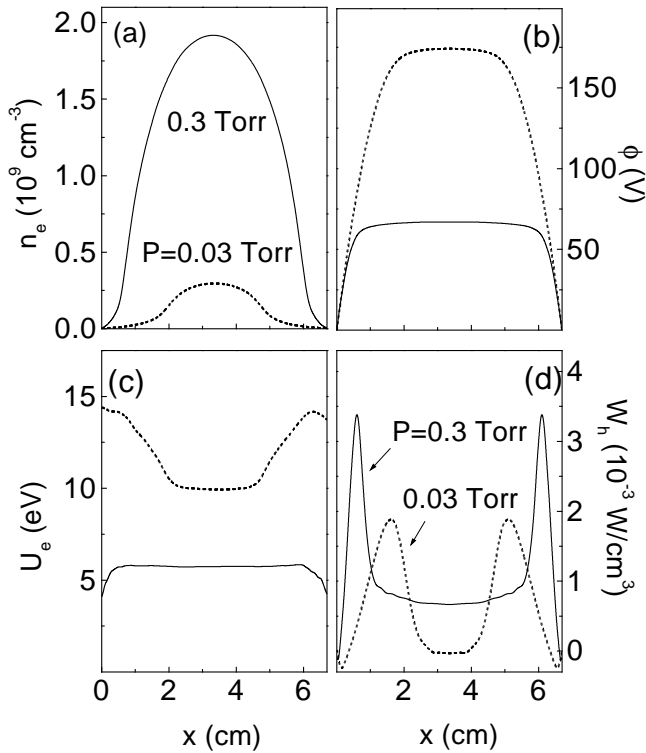}}
\caption {
Spatial distribution of averaged over period electron 
density (a),
potential of electric field (b), mean electron energy (c), and 
electron
heating rate (d) in helium for two gas pressures
$P=0.03$~(dashed lines) and $0.3$~Torr (solid lines),
$d=6.7$~cm, $j=1$~mA/cm$^2$ and  $N=5000$.
}
\label
{dis_helium}
\end{figure}
\begin{figure}[tbp]
\resizebox{1.\columnwidth}{!}
  {\includegraphics[draft=false]{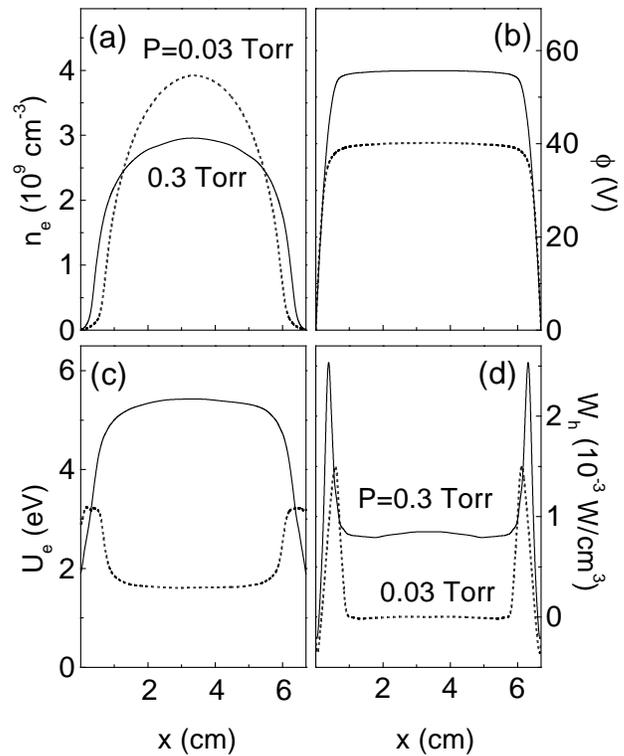}}
\caption {
Spatial distribution of averaged over period electron 
density (a),
potential of electric field (b), mean electron energy (c), and 
electron
heating rate (d) in argon for two  gas pressures
$P=0.03$~(dashed lines) and $0.3$~Torr (solid lines), 
$d=6.7$~cm, $j=1$~mA/cm$^2$ and  $N=5000$.
}
\label
{dis_argon}
\end{figure} 
The electron density and the mean electron energy from the 
experiment \cite {Godyak} and from the combined PIC MCC 
simulations with $N=5000$ are shown in Fig.~\ref {density} as a 
function of gas pressure. The calculated dependence of the 
$\epsilon$ from $P$ demonstrates the transition between 
different modes of the electron heating found in \cite {Godyak} 
and well agrees with the experimental data.

\section {Validity of the combined PIC-MCC approach. Simulation 
results of a ccrf discharge in helium and argon.}

Depending on gas pressure, there are two different 
regimes of electron
heating (collision and collisionless) in rf discharges which are 
well studied experimentally and numerically (see, for example  
\cite {bird,Godyak,Levitskii,Parker,Vahedi,Boeuf}). 
The collision electron
heating takes place due to elastic scattering of the electrons 
on the atoms, when the
directed velocity transfers into the thermal one. 
At high gas pressures 
the collisonal (or ohmic) heating controls the electron energy 
 in the quasineutral part of the discharge. At the low gas 
pressure the electrons are heated due to interaction with moving 
sheaths boundaries and the ohmic
heating in bulk is very small. For these two regimes
the spatial distributions of 
 the electron density, the electrical potential,
 the mean electron energy  
and the electron heating 
rate $W_h=-eE\int v_{ex}f_ed\vec v_e$
are shown in 
Figs.~\ref{dis_helium},\ref{dis_argon} in helium and argon, 
respectively. 

The results are obtained
 for two different gas pressures $P=0.03$~Torr and $P=0.3$~Torr, 
for $d=6.7$~cm 
and $j=1$~mA/cm$^2$.
\begin{figure}[tbp]
\resizebox{.7\columnwidth}{!}
  {\includegraphics[draft=false]{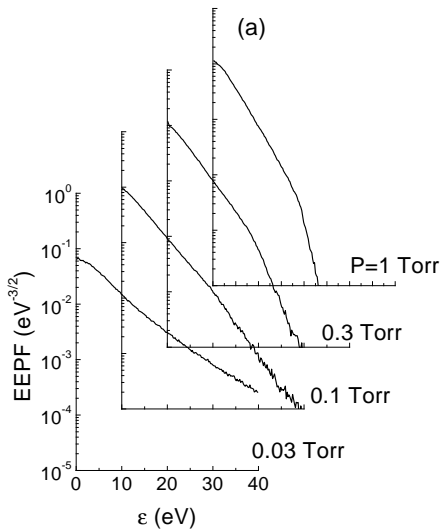}}
\caption {Electron energy probability function in helium  
in the discharge
center ($x=3.35$~cm) for different gas pressures, 
$d=6.7$~cm, $j=1$~mA/cm$^2$ and  $N=5000$.
}
\label
{eepf_helium}
\end{figure}  
\begin{figure}[tbp]
\resizebox{.7\columnwidth}{!}
  {\includegraphics[draft=false]{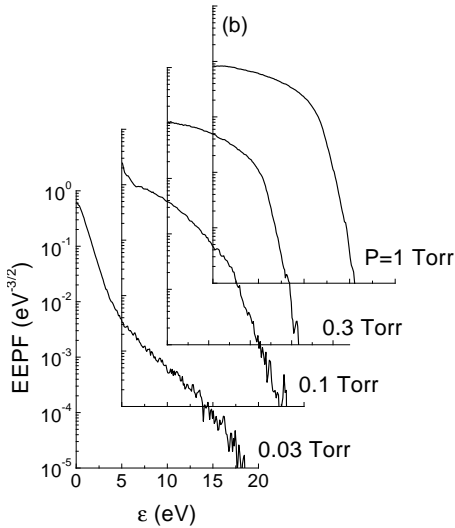}}
\caption {Electron energy probability function in argon  
in the discharge
center ($x=3.35$~cm) for different gas pressures, 
$d=6.7$~cm, $j=1$~mA/cm$^2$ and  $N=5000$.
}
\label
{eepf_argon}
\end{figure}  
As expected, in helium 
 the mean electron energy increases with pressure
lowering in order to compensate an increase of particle losses 
at the electrodes and in argon we observed the opposite 
behavior.
The reasons of
reduction of the electron energy under pressure lowering in 
argon are discussed in 
\cite{tsendin}, where a drop
of the electron temperature up to the gas temperature is 
predicted in the absence of Coulomb electron collisions.
Note, that in helium larger heating rate (in the center of 
discharge) refers to lower $\epsilon$.
This non-local effect can not be predicted within the fluid or
the diffusion-drift approaches. 
 The electron energy probability functions (EEPF) are shown in 
Figs.~\ref {eepf_helium},\ref{eepf_argon}
for helium and argon. 
The data presented in these figures 
averaged over the discharge period. As in the experiment \cite 
{Godyak} we also found in argon that the EEPF changes from a 
Druyvesteyn shape to a bi-Maxwellian one with decreasing the gas 
pressure (see, Fig.~\ref{eepf_argon}). 
At the low gas pressure the electrons are separated 
into two groups. The 
cold electrons are not able
to reach the sheath boundary and their ohmic heating is very 
weak
due to Ramsauer minimum in the elastic cross section (see, 
Fig.~\ref{cross}). \begin{figure}[tbp]
\resizebox{.6\columnwidth}{!}
  {\includegraphics[draft=false]{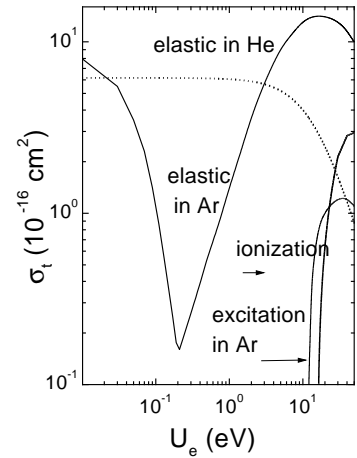}}
\caption {
Electron-neutral elastic cross sections in argon (solid line) 
and in helium (dotted line)
as functions of the electron energy.
}
\label
{cross}
\end{figure}
The fast electrons heated in the sheaths 
maintain the discharge
operation and provide the gas ionization. 
Fig.~\ref {electron_energy}
presents the computed and measured \cite{Godyak} electron 
temperature ($T_e=2U_e/3$) in the discharge
center ($x=3.35~cm$). 
\begin{figure}[tbp]
\resizebox{.7\columnwidth}{!}
  {\includegraphics[draft=false]{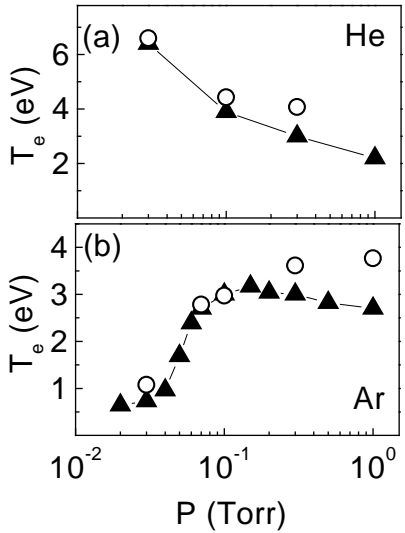}}
\caption {Effective electron temperature 
($T_e=2U_e/3$) in the
discharge center ($x=3.35$~cm) in  helium (a) and argon (b). 
Computed $T_e$ (circles) and measured $T_e$ 
\protect\cite {Godyak} 
(triangles) for  
$d=6.7$~cm, $j=1$~A/cm$^2$ and  $N=5000$.
}
\label
{electron_energy}
\end{figure}
The decrease of the gas pressure is 
accompanied with a drop of $\epsilon$. A comparison with 
experimental data shows a good agreement (within $20\div 30\%
$) within a
pressure range $P=0.03\div 0.3$~Torr for helium and for argon.

 The calculation
gives higher energy at the larger gas pressure $P=1$~Torr. 
The difference between computed and measured data at higher gas 
pressure is likely due to
the contribution of metastable states in the ionization 
kinetics, especially
in helium (see, for example \cite {Parker}). 
In the model of electron-neutral collisions  in our simulations 
we do not take into account the multi-step ionization. At low 
gas pressures we have better agreement because the metastable
atoms are deactivated on electrodes and 
 the influence of multi-step ionization reduces.
 The study of ionization kinetics in noble gases is
out of the scope of this article. Note that in our earlier
study \protect\cite{schweig} of the ccrf discharge 
in helium we have 
considered
the metastable atoms and obtained a good agreement with 
experimental data for high
gas pressures. 

In conclusion we have presented the combined PIC-MCC approach 
for fast simulation of the rf discharge over a wide range of gas 
pressures and current densities. 
The validity of the new approach is justified by comparison with 
the experiment data. The advantage of our approach is the 
considerable decrease of the number of simulation particles $N$.
We are able
to reach a speed-up factor of ten for the collision regime and 
even more for the collisionless regime compared with the 
standard PIC-MCC calculations.      
 
This work is supported 
by the NATO Science for Peace Program, Grant No. 974354.

\bibliographystyle{aipproc}   


\IfFileExists{\jobname.bbl}{}
 {\typeout{}
  \typeout{******************************************}
  \typeout{** Please run "bibtex \jobname" to optain}
  \typeout{** the bibliography and then re-run LaTeX}
  \typeout{** twice to fix the references!}
  \typeout{******************************************}
  \typeout{}
 }

\end{document}